\documentclass[conference]{IEEEtran}
\IEEEoverridecommandlockouts
\usepackage{cite}
\usepackage{amsmath,amssymb,amsfonts}
\usepackage{algorithm}
\usepackage{algorithmic}
\usepackage{graphicx}
\usepackage{textcomp}
\usepackage{xcolor}
\def\BibTeX{{\rm B\kern-.05em{\sc i\kern-.025em b}\kern-.08em
    T\kern-.1667em\lower.7ex\hbox{E}\kern-.125emX}}
\def\*#1{\mathbf{#1}}
\def\~#1{\boldsymbol{#1}}
\begin{document}

\title{Two-stage Modeling for Prediction with Confidence
}

\author{
\IEEEauthorblockN{Dangxing Chen}
\IEEEauthorblockA{ Zu Chongzhi Center for Mathematics and Computational Sciences\\
Duke Kunshan University\\
Kunshan, Jiangsu, China\\
dangxing.chen@dukekunshan.edu.cn}
}

\maketitle

\begin{abstract}
The use of neural networks has been very successful in a wide variety of applications. However, it has recently been observed that it is difficult to generalize the performance of neural networks under the condition of distributional shift. Several efforts have been made to identify potential out-of-distribution inputs. Although existing literature has made significant progress with regard to images and textual data, finance has been overlooked. The aim of this paper is to investigate the distribution shift in the credit scoring problem, one of the most important applications of finance. For the potential distribution shift problem, we propose a novel two-stage model. Using the out-of-distribution detection method, data is first separated into confident and unconfident sets. As a second step, we utilize the domain knowledge with a mean-variance optimization in order to provide reliable bounds for unconfident samples. Using empirical results, we demonstrate that our model offers reliable predictions for the vast majority of datasets. It is only a small portion of the dataset that is inherently difficult to judge, and we leave them to the judgment of human beings. Based on the two-stage model, highly confident predictions have been made and potential risks associated with the model have been significantly reduced. 
\end{abstract}


\section{Introduction}

The use of deep neural networks (DNNs) has been widely successful in many applications, including computer vision \cite{he2016deep}, natural language processing \cite{radford2019language}, speech recognition \cite{hinton2012deep}, bioinformatics \cite{alipanahi2015predicting}, and self-driving cars \cite{bojarski2016end}. Predictive power is not the only factor that is critical for high-stakes applications. Accurate estimates of predictive uncertainty are equally important. In particular, the network should be able to \textbf{know what it does not know}. A self-driving car, for example, can alert its human operators when it detects a situation in which it is unable to respond appropriately and thus control the risk. The distribution of the population can differ significantly from the distribution of the training set in practice. DNNs are frequently challenged by such distribution shifts and out-of-distribution inputs. Consequently, DNNs may become overconfident about their predictions if predictive uncertainty is not quantified, which may result in catastrophic outcomes. 

There has been tremendous research attention paid to this challenge in recent years \cite{lakshminarayanan2017simple, kardan2021towards, bibas2021single, liang2017enhancing}. There are some surveys available at \cite{yang2021generalized,geng2020recent,zhou2021domain}. In \cite{ovadia2019can}, there is a detailed comparison of existing methods. While they have had considerable success, most of the work has been focused on the analysis of images and texts. The purpose of this paper is to examine problems related to finance, another area that is highly regulated. To illustrate our methodology, we analyze the credit scoring problem, but the methodology can be generalized to other problems in finance and applied to a variety of other areas. The field of credit assessment has always been important in finance, and has become even more important with the rise of decentralized finance (DeFi). As of 2020, the blockchain leading protocol Teller raised one million dollars to provide credit scores for Defi crypto loans \footnote{https://decrypt.co/35950/teller-raises-one-million-credit-scores-defi-crypto-loans}. The Credit DeFi Alliance (CreDA) launched a credit rating service in 2021, which used data from multiple blockchains in order to determine a user's creditworthiness \footnote{https://coinrivet.com/new-defi-platform-creda-aims-to-de-risk-the-world-of-crypto/}. As a result, several attempts have been made to develop a reliable modern credit scoring system \cite {dumitrescu2022machine, finlay2011multiple, lessmann2015benchmarking}. However, in spite of their success in improving model performance, predictive uncertainty has been neglected in the existing literature. 

In the credit scoring problem, there is a natural distributional shift. The available training dataset only includes applications that have been approved previously. In the case of institution lending, companies employ certain models, such as logistic regression, in order to exclude potential high-risk clients; in the case of peer-to-peer lending (P2P), individuals make their own judgments in selecting favored loans. If we apply machine learning (ML) methods, trained by approved applicants, directly to open-world datasets, predictive accuracy cannot be guaranteed. In the management of model risk, robustness to various inputs is an important consideration. The Office of the Comptroller of the Currency (OCC) has published a handbook on model risk management \cite{OCC2021model} that stresses the importance of verifying model performance across a variety of inputs. A model can be used safely only if it can provide a degree of confidence in each prediction.  

As a result of the distribution shift challenge, we propose a two-stage model that can provide confident predictions. In the first stage, we use the OOD detection method proposed in \cite{lakshminarayanan2017simple} to quantify the predictive uncertainty. To determine how volatile the prediction is at each input, the ensemble method with random initializations and random shuffling of neural networks (NNs) is applied. It is safe to predict with a mean of NNs if the variance of the prediction to a given input is small; inputs with high variance are classified as unconfident and should not be directly predicted by the model. The second stage involves utilizing the domain knowledge of the problem to further investigate the unconfident set. Monotonic relationships exist between the probability of default (PoD) and various features of credit scoring. As an example, a customer with more past dues is more likely to default. In light of this knowledge, we propose a mean-variance optimization strategy to provide lower bounds for unconfident data points. To be more precise, we are looking for a data point with a high PoD and a low variance that is close to an unconfident point. When the PoD of the founded point has passed the decision threshold, predictions can be made with confidence. Unconfident points with unsatisfactory lower bounds are classified as undecided and are left for human decision. Although our two-stage modeling could not provide accurate estimations for unconfident sets, lower bounds usually provide sufficient information to make informed decisions. 

The power of our method is demonstrated through the use of a representative empirical example. First, we demonstrate that the OOD detection method can be successfully used to identify OOD data by carrying out an auxiliary experiment mimicking a real-world problem. The study also demonstrates that NNs are not confident about a small portion of the dataset, and that predictions should not be made directly based on that portion of the dataset. We then produce lower bounds for the data points in the unconfident set in the second stage. By using estimated lower bounds and a mild threshold, we are able to make decisions for most samples in the unconfident set. 
This results in only a very small portion of the dataset that cannot be decided, and these are left to the undecided set. In addition, we examine a case from the undecided set. It has been found that some samples are inherently difficult to judge, and therefore they should be left for human investigation. In summary, our two-stage model has produced confidence predictions for the majority of our dataset with great success. 

The remainder of the paper is organized as follows. In Section 2, we present our two-stage model. Section 3 illustrates an empirical experiment. Section 4 concludes our discussion.

\section{Two-stage modeling}

Assume we have $\mathcal{D} \times \mathcal{Y}$, where $\mathcal{D}$ is the dataset with $n$ samples and $p$ features and $\mathcal{Y}$ represents corresponding classification labels. We consider the binary classification such that
\begin{align*}
y_i = \begin{cases}
1, & \text{default}, \\
0, & \text{not default}. 
\end{cases}
\end{align*}
We are interested in the probability of default (PoD) of applicants given their information. A credit score is then calculated based on the PoD. We assume that
\begin{align*}
\mathbb{E}[y|\*x] = g(f(\*x)) 
\end{align*}
for some continuous function $f$, and $
g(x) = \frac{1}{1+e^{-x}}$.
Machine learning (ML) models are then applied to learn $f$. Here, we consider neural networks (NNs) for their universal approximation property. 

Real-world samples of $\*x$ are drawn from some unknown universal distribution $h(\*x)$. It is expected that the model will have good generalization if the population has the same distribution. The situation is different for credit scoring problems, however. As for institution lending, companies use certain criteria, such as logistic regression, to determine which applicants are likely to default. As for peer-to-peer (P2P) lending, loaners use their own judgment in making loan decisions. Consequently, we are unable to observe all data. Specifically, we are unaware of the outcome of the data that indicates that the applicant has been rejected. This may negatively affect the performance of ML methods, as there are some regions of $\*x$ that models have not been exposed to during the training processes.

A key aspect of model risk management is checking the robustness of the model and understanding its scope. In \cite{OCC2021model}, it stated, ``If testing indicates that the model may be inaccurate or unstable in some circumstances, management should consider modifying certain model properties, putting less reliance on its outputs, placing limits on model use, or developing a new approach." In light of this, we have developed a two-stage system. For the first stage, we train an ensemble of neural networks for the prediction with confidence. As a result, OOD data is detected when a certain threshold is met. Using the domain knowledge of the problem, we solve a mean-variance optimization to determine the lower bound of the PoD for OOD data. Therefore, even though accurate estimates of PoD of OOD data are not available, lower bounds are often sufficient for decision making.

\subsection{Out-of-distribution detection}

We follow \cite{lakshminarayanan2017simple} as a simple yet effective approach to detect OOD data. Here, we are concerned with the level of confidence that a model has regarding a particular data point. In the past, ensemble methods have been used to improve the performance of models. \cite{lakshminarayanan2017simple} has recently shown that ensemble methods can also be used to estimate uncertainty. 
NNs are trained using randomized initialization, which results in different NNs each time, due to the lack of convexity of the loss function. Hence, by examining the behavior of NNs for each sample point, we are able to determine how uncertain the predictions of NNs. The entire training dataset is used to train NNs for better performance over bagging, as discussed in \cite{lakshminarayanan2017simple} and \cite{lee2015m}. In each training, NN parameters are randomly initialized and the data points are randomly shuffled in order to train a collection of NNs $\{ f_i(\*x; \~{\theta}_i) \}$. We may then be able to make predictions based on their means
\begin{align}
    \widehat{\mu}(\*x) = \frac{1}{m} \sum_{i=1}^m g(f_i(\*x; \~{\theta}_i)),
\end{align}
where $m$ is the number of NNs. 
Variance is used to determine the volatility of a prediction
\begin{align}
    \widehat{\sigma}^2(\*x) = \frac{1}{m-1} \sum_{i=1}^m (g(f_i(\*x; \~{\theta}_i)) - \widehat{\mu}(\*x))^2.
\end{align}
It is not recommended to use $\widehat{\mu}(\*x)$ if $\widehat{\sigma}^2(\*x)$ is more than certain thresholds $\epsilon$. Correspondingly, we split the dataset into the confident set and unconfident set.

\subsection{Mean-variance optimization}

In the case of $\*x$ where $\widehat{\sigma}^2(\*x)>\epsilon$, we are unable to trust the model for prediction from the data perspective. It is possible, however, to make a prediction using domain knowledge. Several features of credit scoring have monotonic relationships with respect to their outputs. For instance, the PoD should increase monotonically with respect to delinquency features: customers who have more past dues are more likely to default.

Let us assume, for simplicity's sake, that the output is monotonically increasing with respect to the desired features. For monotonic decreasing features, we can simply add a negative sign.  Suppose $\~{\alpha}$ is the subset of all monotonic features and $\neg \~{\alpha}$ its complement, then the input $\*x$ can be partitioned into $\*x = (\*x_{\~{\alpha}}, \*x_{\neg \alpha})$. Suppose $\mathcal{X}$, $\mathcal{X}_{\~{\alpha}}$, $\mathcal{X}_{\neg \~{\alpha}}$ are spaces of $\*x, \*x_{\~{\alpha}}, \*x_{\neg \~{\alpha}}$, respectively. We say $f$ is monotonic with respect to $\*x_{\~{\alpha}}$ if 
\begin{align*}
& f(\*x_{\~{\alpha}}, \*x_{\neg \~{\alpha}}) \leq f(\*x'_{\~{\alpha}}, \*x_{\neg \~{\alpha}}),  \\
& \forall \*x_{\~{\alpha}} \leq \*x'_{\~{\alpha}}, \forall \*x_{\~{\alpha}}, \*x_{\~{\alpha}}' \in \mathcal{X}, \forall \*x_{\neg \~{\alpha}} \in \mathcal{X}_{\neg \~{\alpha}},
\end{align*}
where $\*x_{\~{\alpha}} \leq \*x_{\~{\alpha}}'$ denotes the inequality for all entries, i.e., $x_{i} \leq x_{i}'$ for all $i \in \alpha$.

\begin{figure}
    \centering
    \includegraphics[scale=0.32]{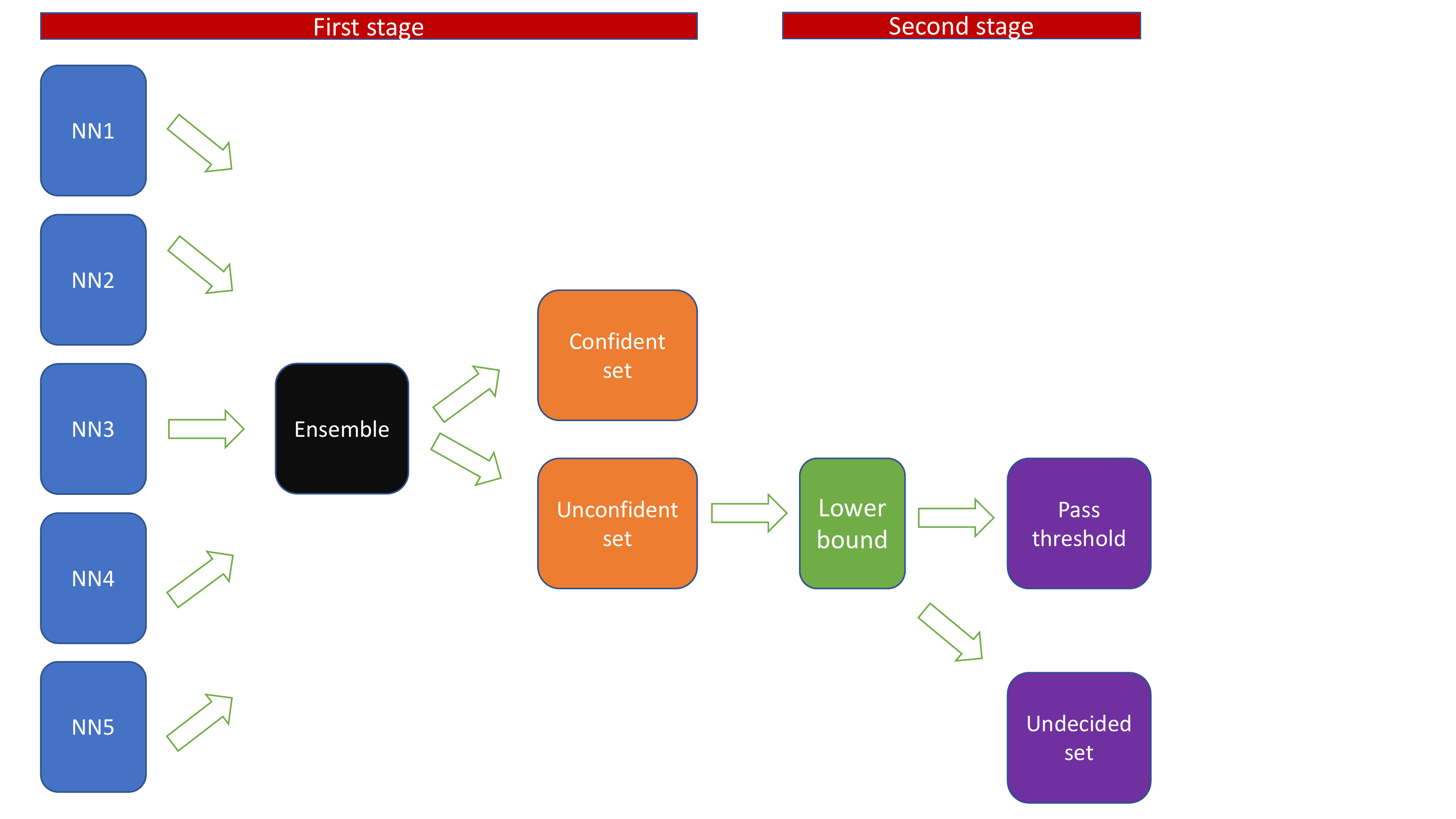}
    \caption{Two-stage model}
    \label{fig:two-stage}
\end{figure}

The monotonicity could be used to provide some information. For $\*x = (\*x_{\~{\alpha}}, \*x_{\neg \~{\alpha}})$, suppose we could identify another point $\*x' = (\*x'_{\~{\alpha}}, \*x_{\neg \~{\alpha}})$ that is close to $\*x$, with $\*x'_{\~{\alpha}} \leq  \*x_{\~{\alpha}}$ or $\*x'_{\~{\alpha}} \geq  \*x_{\~{\alpha}}$ and we are confident in prediction in $\*x'$, then we would be able to provide a reliable lower and upper bound for $\*x$. For simplicity, we will focus on the lower bound case, and the upper bound case will be similar. The ideal candidate $\*x'$ has a large value of $\widehat{\mu}(\*x')$ and a small value of $\widehat{\sigma}^2(\*x')$. Thus, we have a high probability that the applicant will default, and we are quite confident about this outcome. It is necessary to perform an optimization in order to determine such a point. However, it is not possible to optimize both mean and variance at the same time. As a result, we propose a constraint optimization mean-variance strategy, which is commonly used in modern portfolio theory for risk management \cite{kolm201460}
\begin{align} \label{eq:mean-variance}
\begin{cases}
    \max_{\*x'_{\~{\alpha}}} \ & \widehat{\mu}(\*x'), \\
    \text{subject to } & \widehat{\sigma}^2(\*x') < \epsilon, \\
    & \*x'_{\~{\alpha}} \leq  \*x_{\~{\alpha}}, \\
    & \*x'_{\~{\alpha}} \in \mathcal{X}_{\~{\alpha}}.
\end{cases}
\end{align}
The constraint optimization \eqref{eq:mean-variance} can be easily solved by many existing packages. 
By identifying $\*x'$, although an accurate estimation is not available, one might be able to make the decision with confidence if $\widehat{\mu}(\*x')$ is sufficiently large. The remaining unconfident set is then split into two sets: the set that passes the thresholds with $\widehat{\mu}(\*x')>\tau$ and $\widehat{\sigma}(\*x')<\epsilon$, enabling a decision to be made, and another undecided set that the model is unable to predict, which is left for further analysis by humans. A graphical summary of the two-stage model is presented in Figure~\ref{fig:two-stage}.

It is possible to calculate  upper bounds in a similar manner, but we do not do so here. The reason for this is that for credit scoring, OOD samples are usually found in large values of features. For instance, applicants with zero past dues are common in the dataset, whereas applicants with many past dues are unlikely to have been approved previously and therefore are rare. As a result of calculating upper bounds, optimization is ill-conditioned because they explore a large region where a significant portion is outside the domain of distribution.

\section{Empirical example}

In this study, we focus on a popularly used  Kaggle credit score dataset \footnote{https://www.kaggle.com/c/GiveMeSomeCredit/overview}. As other datasets have similar results, we have omitted them from this analysis. We have removed data with missing variables in order to simplify the analysis. In this paper, careful data cleaning is not the primary focus, although it is possible to improve model performance if the data is cleaned carefully. The dataset is randomly divided into $75\%$ training and $25\%$ test sets. Ten features are included in the dataset as explanatory variables:
\begin{itemize}
\item $x_1$: Total balance on credit cards and personal lines of credit except for real estate and no installment debt such as car loans divided by the sum of credit limits (percentage).
\item $x_2$: Age of borrower in years (integer).
\item $x_3$: Number of times borrower has been 30-59 days past due but no worse in the last 2 years (integer).
\item $x_4$: Monthly debt payments, alimony, and living costs divided by monthly gross income (percentage). 
\item $x_5$: Monthly income (real).
\item $x_6$: Number of open loans (installments such as car loan or mortgage) and lines of credit (e.g., credit cards) (integer). 
\item $x_7$: Number of times borrower has been 90 days or more past due (integer). 
\item $x_8$: Number of mortgage and real estate loans including home equity lines of credit (integer). 
\item $x_9$: Number of times borrower has been 60-89 days past due but no worse in the last 2 years (integer). 
\item $x_{10}$: Number of dependents in the family, excluding themselves (spouse, children, etc.) (integer). 
\item $y$: Client's behavior; 1 = Person experienced 90 days past due delinquency or worse.
\end{itemize}

In terms of model architecture, NNs contain one hidden layer with two units, logistic activation, and no regulation. Our results indicate that more complex architectures do not necessarily enhance model performance. As a result, we continue to follow this simple structure to avoid overfitting. 
We perform a two-stage analysis in this study. In the first stage, we examine whether the out-of-distribution (OOD) detection is effective by considering an auxiliary experiment, and then we investigate the entire dataset. During the second stage, we apply the mean-variance optimization to provide information for samples that our model is not confident about.

\subsection{Out-of-distribution detection}

Our objective is to test whether our method can detect OOD data. Unfortunately, the entire population of the dataset is not available, and the publicly available dataset only contains data related to approved applicants. In practice, loans are approved in accordance with certain criteria. Institutional lenders use models, such as logistic regression, to exclude high-risk clients; peer-to-peer lenders (P2P) select loans using their own judgment. Because of this, only a portion of the dataset is selected for training, and the trained model is not necessarily accurate for data that has not been seen. 
In order to mimic the data selection process, we propose the following auxiliary experiment. Denote the training set as $\mathcal{D}_1$ and the test set as $\mathcal{D}_2$. For $\mathcal{D}_1$, a neural network $f_1$ is used to train in order to determine a decision boundary. Then, $f_1$ further splits $\mathcal{D}_1$ and $\mathcal{D}_2$ into selected sets $\mathcal{D}_{1,1}, \mathcal{D}_{2,1}$ and unselected sets $\mathcal{D}_{1,2}, \mathcal{D}_{2,2}$ with a threshold $\tau=0.5$. Then we train our model $f_2$ to the selected set $\mathcal{D}_{1,1}$ and exclude $\mathcal{D}_{1,2}$. By checking whether samples of $\mathcal{D}_{2,2}$ can be identified by OOD detection, we can determine the performance of the method. 

During the experiment, we used $m=10$ and $\epsilon=10^{-3}$. As a result of fitting $f_1$ to $\mathcal{D}_1$, $2.1\%$ of the dataset is predicted as default. In the test set, $2.0\%$ of the dataset belongs to $\mathcal{D}_{2,2}$. Then we fit $f_2$ to $\mathcal{D}_{1,1}$ and apply it to $\mathcal{D}_2$. The averaged area under the curve (AUC) in $\mathcal{D}_2$ is $81.0\%$, indicating its accuracy. In the case of $f_2$, there is uncertainty about $5.4\%$ of the test set. Among the unconfident set, it covers $96.2\%$ samples in $\mathcal{D}_{2,2}$, demonstrating its power. If we reduce $\epsilon$ by a factor of 2, it covers all samples, indicating that some samples in $\mathcal{D}_{2,2}$ are very close to the distribution boundary. Considering the fact that $f_1$ is not confident for all samples, it is not surprising that $f_2$ is unconfident with more samples than the size of $\mathcal{D}_{2,2}$. Based on the results of this experiment, the method for detecting OOD appears to be successful. 

\begin{figure}
    \centering
    \includegraphics[scale=0.45]{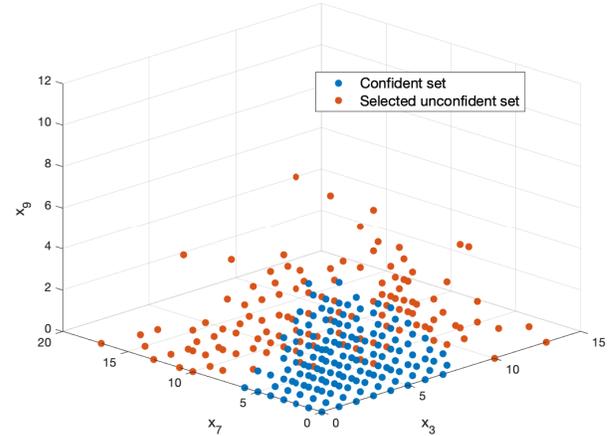}
    \caption{Out-of-distribution visualization}
    \label{fig:cluster_379}
\end{figure}

For a more comprehensive investigation, we have now applied OOD detection to the entire dataset. As a result, $3\%$ of the dataset is determined as the unconfident. This is somewhat lower than the $5.4\%$ in the previous example, as we have included more data for training. 
We wish to understand what data has been classified as unconfident. In order to reduce the dimensionality of the ensembled NNs, we use the sensitivity-based analysis \cite{horel2018} for feature importance.
As a result, $x_3$, $x_7$, and $x_9$ have explained over $90\%$ of importance collectively. In order to visualize the results, we focus on these three features. In Figure~\ref{fig:cluster_379}, we plot samples whose summation of these three features exceeds ten but less than twenty (for better visualization) and compare them with samples in the confidence set. Indeed, these samples are either outside of the domain or close to it. 


\subsection{Mean-variance optimization}

In the second stage, the mean-variance optimization is then applied to these data points. The first step in this process is to determine monotonic features. Based on domain knowledge, we assume that the output will increase monotonically over $x_3$, $x_5$, $x_7$, $x_9$, and $x_{10}$. We obtain lower bounds by solving \eqref{eq:mean-variance}. Depending on the choice of $\tau$, the unconfident set is divided into two groups. It is true that $\tau=50\%$ is a natural criterion for classification problems, however, for credit scoring, companies often choose a lower criterion in order to reduce the risk associated with their loans. It is important to note that $50\%$ represents a very high probability of default. There are various choices of $\tau$ depending on the companies' risk appetite. Here we calculate the undecided portion based on the different choices of $\tau$ in Table~\ref{tab:GMSC_OOD}. We will be able to make more decisions if we make a more conservative choice. With $\tau=10\%$, decisions can be made with confidence about $90\%$ of the unconfident set, which only leaves $0.3\%$ samples out of the entire dataset in the undecided set. We have utilized most of the dataset in such cases, with only a small portion considered to be undecided.

\begin{table}[h]
    \centering
    \caption{Model Performance of the Taiwan dataset.}
    \begin{tabular}{ccc}
    \hline
    Threshold $\tau$  & Undecided portion  \\ \hline
    $50\%$ & $89.0\%$ \\ \hline
    $40\%$ & $74.2\%$ \\ \hline
    $30\%$ & $33.4\%$  \\ \hline
    $20\%$ & $15.4\%$  \\ \hline
    $10\%$ & $9.6\%$ \\ \hline 
    \\
    \end{tabular}
    \label{tab:GMSC_OOD}
\end{table}

To gain a deeper understanding of the results, we take a close look at them. As an example, one sample in the unconfident set is
\begin{align*}
    \*x_a = \left[ \begin{matrix} 2.2 & 34 & 1 & 0.095 & 3000 & 2 & 15 & 0 & 0 & 0 \end{matrix} \right].
\end{align*}
What is unusual about $\*x_a$ is that there are 15 times that the borrower has been 90 days or more past due, whereas the average number is 0.21. Additionally, the borrower has only one past due between 30 and 59 days. It is quite unusual to see such behavior in the dataset, and it is quite rare. Due to this, it has a prediction variance of $1.5  \times 10^{-3}$, and we do not feel confident about its accuracy. Using the mean-variance optimization, we find that by modifying $x_7$ from 15 to 5, we can reduce the variance to $1 \times 10^{-3}$, which is below the threshold of $\epsilon$. As we have more data with a small amount of past dues, this intuitively makes sense as well. Meanwhile, the new PoD is about $61.2\%$, which is already high. As a result, we are able to safely reject the applicant based on $\*x_a$. 

In addition, we examine another example that leads to the undecided set. The following is an example of a sample 
\begin{align*}
    \*x_b = \left[ \begin{matrix} 0.003 & 53 & 0 & 6.0 & 8000 & 58 & 0 & 54 & 0 & 0 \end{matrix} \right].
\end{align*}
In this case, the optimization did not find a minimizer that satisfied the constraints. There is no delinquency record for this applicant. There are, however, a number of outstanding loans, lines of credit, mortgages, and real estate loans. In the dataset, the sample in question is extremely rare and, as a result, is highly suspicious. Due to the rarity of this sample, the area surrounding it is out of bounds. Our optimization algorithms fail because we did not place more restrictions on these features. With $x_6$ and $x_8$ set to 0, the prediction variance becomes $1.4 \times 10^{-5}$, which is highly reliable. Nevertheless, there do not exist natural monotonic relationships globally between these two features and the output. It may be possible to address this problem by imposing some local constraints. However, this requires a greater degree of domain knowledge, which is omitted here for brevity. Therefore, we cannot rely on domain knowledge to provide reliable bounds. It would be beneficial to inquire further about the circumstances of this particular case by asking the customer for more details. As an example, why do the customers have so many loans? In the event that we are not able to find a good lower bound, we should return it with an error message for human decision-making.

\section{Conclusion}

In this paper, we propose a two-stage model which provides a highly confident prediction that significantly reduces the model risk. Using the out-of-distribution detection method, data is divided into confident and unconfident sets in the first stage. In the second stage, the unconfident data is further examined. Lower bounds can be provided by solving mean-variance optimizations, allowing decisions to be made for a large portion of unconfident data. Only a small portion of the dataset is left for human judgment. By utilizing domain knowledge, we are able to provide information for samples that have not been analyzed previously by models. This study focuses on credit scoring. In the future, we will extend our framework to other areas with domain expertise.

\bibliographystyle{IEEEtran}
\bibliography{chen}

\end{document}